\documentclass[conference, twocolumn, 10pt]{IEEEtran}

\ifCLASSINFOpdf
  \usepackage[pdftex]{graphicx}
   \graphicspath{{..1/pdf/}{../jpeg/}}
   \DeclareGraphicsExtensions{.pdf,.jpeg,.png}
\else
   \usepackage[dvips]{graphicx}
   \graphicspath{{../eps/}}
   \DeclareGraphicsExtensions{.eps}
\fi

\usepackage{cite}
\usepackage{comment} 
\usepackage[cmex10]{amsmath}
\usepackage{amsfonts, amsmath, amsthm, amstext, mathrsfs, graphicx, color, url}
\usepackage{array}
\usepackage{mdwmath}
\usepackage{mdwtab}
\usepackage{algpseudocode}
\usepackage{algorithm}
\usepackage{url}
\usepackage{setspace}
\usepackage{footnote}
\usepackage[skip=0pt]{caption}
\usepackage{float}
\usepackage{epstopdf}
\usepackage{epsfig, amsmath,amssymb,epsf,cite,subfigure,scalefnt,multirow,array,setspace}
\newcommand\subparagraph{%
  \@startsection{subparagraph}{5}
  {\parindent}
  {3.25ex \@plus 1ex \@minus .2ex}
  {-1em}
  {\normalfont\normalsize\bfseries}}
\makeatother
\usepackage{titlesec}
\let\subparagraph\relax 
\titlespacing*{\section}
{0pt}{2pt}{2pt}
\titlespacing*{\subsection}
{0pt}{2pt}{2pt}
\titlespacing*{\subsubsection}
{0pt}{2pt}{2pt}

\def\IC{{\mathbb C}}

\def\calI{{\mathcal I}}

\def\calK{{\mathcal K}}
\def\calL{{\mathcal L}}

\def\calO{{\mathcal O}}

\def\bA{{\pmb A}}

\def\bH{{\pmb H}}
\def\bI{{\pmb I}}

\def\bK{{\pmb K}}
\def\bL{{\pmb L}}
\def\bM{{\pmb M}}

\def\bQ{{\pmb Q}}
\def\bR{{\pmb R}}

\def\bU{{\pmb U}}
\def\bV{{\pmb V}}

\def\bX{{\pmb X}}

\def\bf{{\pmb f}}

\def\bs{{\pmb s}}

\newcommand{\bPsi}{  \pmb{\Psi}  }

\newcommand{\bLam}{  \pmb{\Lambda}  }
\newcommand{\bTh}{  \pmb{\Theta}  }

\newcommand{\upperRomannumeral}[1]{\uppercase\expandafter{\romannumeral#1}}

\newcommand{\st}{ \textup{s. t.} }

\newcommand{\argmax}{ \textup{argmax} }

\def\bSig{ \mathbf{\Sigma} }

\def\tr{{\textup{tr}}}

\newcommand{\lrb}[1]{ \lbrace #1 \rbrace }

\newtheorem{lemma}{Lemma}

\newtheorem{proposition}{Proposition}

\begin{document}

\title{Low-Overhead Coordination in \\ Sub-28 Millimeter-Wave Networks  }

\author{\IEEEauthorblockN{Hadi Ghauch}
\IEEEauthorblockA{School of Electrical Engrng \\
Royal Inst. of Techn., KTH\\
ghauch@kth.se }
\and
\IEEEauthorblockN{Taejoon Kim}
\IEEEauthorblockA{Dept. of Electrical Engrng\\ and Computer Science\\
Univ of Kansas, Lawrence\\
taejoonkim@ku.edu }
\and 
\IEEEauthorblockN{Mikael Skoglund}
\IEEEauthorblockA{School of Electrical Engineering\\
Royal Inst. of Techn., KTH \\
skoglund@kth.se }
\and 
\IEEEauthorblockN{Carlo Fischione}
\IEEEauthorblockA{School of Electrical Engrng\\
Royal Inst. of Techn., KTH \\
carlofi@kth.se }
}
\maketitle

\begin{abstract}
In this paper, we present some contributions from our recent investigation. 
We address the open issue of interference coordination for sub-$28$ GHz millimeter-wave communication, by proposing fast-converging coordination algorithms, for dense multi-user multi-cell networks. 
We propose to optimize a lower bound on the network sum-rate, after investigating its tightness.  
The bound in question results in distributed optimization, requiring local information at each base station and user. We derive the optimal solution to the transmit and receive filter updates, that we dub non-homogeneous waterfilling, and show its convergence to a stationary point of the bound.  
We also underline a built-in mechanism to turn-off data streams with low SINR, and allocate power to high-SNR streams.   
This ``stream control'' is a at the root of the fast-converging nature of the algorithm.  
Our numerical result conclude that low-overhead coordination offers large gains, for dense sub-$28$ GHz systems. These findings bear direct relevance to the ongoing discussions around 5G 
New~Radio.  
\end{abstract}
\begin{IEEEkeywords}
Sub-$28$ GHz Millimeter-wave, low-overhead coordination, Difference of Log and Trace (DLT), Non-homogeneous Waterfilling, max-DLT. 
\end{IEEEkeywords}
\IEEEpeerreviewmaketitle

\section{Introduction}
To address the exponentially increasing demand in 5G systems, communications in the millimeter-wave (mmWave) band are among the most promising candidates~\cite{Andrews_5G_14}, due to the large mmWave spectrum. While most investigations of mmWave communcation have been focused on systems above $28$ GHz, in the current work, we study multi-user multi-cell coordination, in \emph{sub-$28$ GHz systems}, e.g., X-band ($8$-$12$) GHz, Ku-band ($12$-$18$) GHz, and $28$ GHz in the Ka band. 
These systems are characterized by a relatively large antenna spacing (compared to systems beyond $60$ GHz), thereby implying that tens (rather than hundreds) of antennas can be fitted on transmitters/receivers. Thus, the urge for hybrid analog-digital precoding is not stringent and \emph{fully digital} precoding/combining is preferred. Moreover, propagation channels are still dominated by Rayleigh/Rician components in non line-of-sight environments~\cite{Rappaprt_mmWprop}. As a result, conventional pilot-based channel estimation techniques are more efficient than beam alignment/sounding~\cite{Hur_mmWave_13}. 

The implication of highly directional wireless links at $60$ GHz (and above) is that they are almost interference-free.  
However, in sub-$28$ GHz systems, channels are \emph{less sparse} (in terms of eigenmodes), and beamforming has relatively lower directivity than systems in the higher bands. This is attributed to the presence of significant multi-path components, in urban propagation (confirmed by narrowband/wideband channel measurements in the $9.6$ GHz, $11.4$ GHz, and $28.8$ GHz bands~\cite{Viotette_11GHz_88}). 

Consequently, interference may still be a limiting factor in these systems, especially when considering \emph{dense} multi-cell scenarios, where interference management and coordination are still beneficial.  
Coordination in multi-user multi-cell networks generally refers to the exchanges of information among base stations, to increase the network sum-rate. 
While these aspects have been investigated at the MAC layer~\cite{Shokri_mmWMACSurvery_15}, they are still essentially unaddressed at the physical layer. Indeed, the benefits/costs of coordination in sub-$28$ GHz systems is still an open problem, especially in the case of ultra dense networks - believed to be pervasive in future networks~\cite{METISD62}: In these scenarios, ignoring interference for cell-edge users may be a limiting factor on the sum-rate. 

In the context of multi-user multi-cell networks, coordination is done using the framework of \emph{Forward-Backward (F-B) iterations}: this over-the-air training leverages local Channel State Information (CSI) at each Base Station (BS) and user, to iteratively optimize the filter at each BS/user, in a fully distributed manner. This framework has been at the heart of most distributed coordination algorithms, such as interference leakage minimization~\cite{gomadam_distributed_2011}, max-SINR~\cite{gomadam_distributed_2011}, minimum mean-squared error~\cite{schmidt_minimum_2009}, and (weighted) sum-rate maximization~\cite{shi_wmmse_2011}.      

Unfortunately, these conventional schemes suffer from extremely elevated overhead, as they require hundreds/thousands of F-B iterations before convergence~\cite{Schmidt_comparison_13}, where the latter increases with the number of BSs, users and transmit/receive antennas~\cite{Schmidt_comparison_13}.   
Moreover, mmWave systems will have a larger number of BSs/cells per unit-area (due to their inherent short range), and require a much larger number of  BS/user antennas (to mitigate pathloss through array gain), compared to sub-$6$ GHz systems. 
Consequently, this shortcoming severely \emph{impairs} the applicability of conventional coordination, to the systems in question.  
This limitation is reinforced by the \emph{lower coherence time} of mmWave channels. 
Despite bearing direct relevance to conventional sub-$6$ GHz, this major limitation has only been addressed in a few recent works~\cite{Komulainen_EffCSI_13, Nguyen_WSR_14, Brandt_FastConv_15, Ghauch_IWU_15}, and remains essentially unexplored. 

\begin{figure}
  \center
  \includegraphics[ scale=.6]{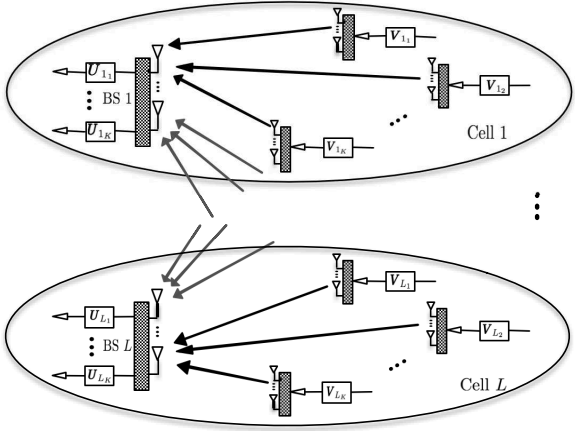}
  \caption{ $L$-cell MIMO Interfering Multiple-Access Channel} 
  \label{fig:setup}
  \vspace{-2em}
\end{figure}
In this work, we design low-overhead distributed coordination algorithms, constrained to operate in a just a few F-B iterations for \emph{increasing dimensions} of the networks. In the algorithm design, we further aim at a \emph{tenfold} reduction in the communication overhead of conventional algorithms. 
We derive and optimize a lower bound the sum-rate maximization problem in MIMO Interfering Multiple-Access Channels (MIMO IMAC), that we dub \emph{Difference of Log and trace (DLT)}. 
Unlike the sum-rate, when combined with alternating optimization methods, the DLT expression results in subproblems that are distributed (only requiring local CSI at each BS and user)~\cite{Ghauch_MAXSEP_16}. Despite their non-convexity, we derive the optimal solution to each of these subproblems, that we dub \emph{non-homogeneous waterfilling} (a variation on the classical waterfilling). 
This solution turns-off data streams with low-SINR, and allocates power to streams with high SNR. The built-in \emph{``stream-control''} is key to achieving the tenfold increase in convergence speed. Moreover, we show that the devised algorithm converges to a locally optimal solution of the DLT bound. 
It is revealed that the proposed fast-converging algorithm offers large sum-rate gains, compared to many standard and fast-converging benchmarks.
Coordination is still a vital aspect of these systems, and offers major performance gains over uncoordinated transmission
While the approach is developed for the MIMO IMAC (i.e., uplink communication), the methods/results are applicable to the downlink, and all its special cases. 

\section{System Model}
The proposed algorithms operate within the framework of F-B iterations/training 
by  exploiting the uplink (UL) and downlink (DL) channel reciprocity. 
The proposed scheme is designed to operate in the low-overhead regime, by restricting the number of F-B iterations, $T$, to $T \leq 5$ (a tenfold reduction in communication overhead, over conventional coordination algorithms).  We assume that each BS and user posses local CSI only, that is assumed to be perfect (i.e., no  CSI errors). 

We consider a multi-user multi-cell setting, with $L$ cells/BSs, serving $K$ users each. In the MIMO IMAC case, transmitters at users and receivers are BSs.    
Each transmitter and receiver have $M$ and $N$ antennas, respectively, and communicate $d$ data streams.  Let $\calL$ be the set of BSs, $\calK_l$ the set of users served by BS $l \in \calL $, and $\calI$ the total set of users.   
Moreover, we denote by user $l_j \in \calI $ the $j$th user ($ j \in \calK_l$), in the $l$th cell ($ l \in \calL$).   
The recovered signal for user $j \in \calK_l $, in cell $ l \in \calL$ (denoted as user $l_j \in \calI $ hereafter),  
\begin{align} 
\pmb{\tilde{s}}_{l_j} &=   \bU_{l_j}^\dagger \bH_{l, l_j} \bV_{l_j} \pmb{s}_{l_j} \nonumber \\
&~~~+ \sum_{\substack{ i \in \calL \\ i \neq l}} \sum_{  k \in \calK }  \bU_{l_j}^\dagger \bH_{l,i_k} \bV_{i_k} \pmb{s}_{i_k} +  \bU_{l_j}^\dagger \pmb{n}_l, \ \forall \ l_j \in \calI
\end{align}
where \footnote{\emph{Notation:} we use bold upper-case letters to denote matrices, and bold lower-case denote vectors. For a given matrix $\pmb{A}$, we define $\tr(\pmb{A})$ as its trace, $\Vert \pmb{A} \Vert_F^2$ as its Frobenius norm, $|\pmb{A}|$ as its determinant,  $\pmb{A}^\dagger$ as its conjugate transpose, and $\bA^{-\dagger}$ as $(\bA^\dagger)^{-1}$.
	In addition, $\bA_{(i)}$ denotes its $i$th column, $\bA_{i:j} $ columns $i$ to $j$,  $\bA_{(i,j)}$ element $(i,j)$ in $\bA$, $\lambda_i[\pmb{A}]$ the $i^{th}$ eigenvalue of a Hermitian matrix $\pmb{A}$ (assuming the eigenvalues are sorted in decreasing order), and $v_{1:d}[\bA] $ denotes the $d$ dominant eigenvectors of $\bA$. 
	Furthermore, $\pmb{A} \succ \pmb{0}$ (resp. $\pmb{A} \succeq \pmb{0}$) implies that $\pmb{A}$ is positive definite (resp. positive semi-definite). 
	Finally, $\pmb{I}_n$ denotes the $n \times n$ identity matrix, $\lrb{n} = \lrb{1, \cdots, n } $, and $x^{+} =  \max \lrb{0, x}$.
}$\bV_{i_k} \in \IC^{M \times d} $ is the linear transmit filter for user $i_k \in \calI$, $\bU_{l_j} \in \IC^{N \times d} $ is the linear receive filter for user $l_j \in \calI$, and $\bH_{l, i_k} $ the $N \times M$ MIMO channel from user $i_k \in \calI$, to BS $l$ assumed to be block-fading.\footnote{The model/results assume $M, N$ and $d$ for simplicity, and can be easily extended to differ across users and BSs.}
$\bs_{i_k} \in \IC^d $ is the transmit signal of user $i_k \in \calI$ with unit power symbols, and  $\pmb{n}_l$ is the AWGM noise at receiver $l$, with $\mathbb{E}[\pmb{n}_l \pmb{n}_l^\dagger ] = \sigma_l^2 \pmb{I}_N $. 

We assume simple decoding, i.e., treating interference as noise , without successive interference cancellation. Then, the achievable rate of user $l_j \in \calI $ is given by, 
\begin{align}
r_{l_j} = \log_2 | \pmb{I}_d + (\bU_{l_j}^\dagger \bR_{l_j} \bU_{l_j}) (\bU_{l_j}^\dagger \bQ_{l_j} \bU_{l_j})^{-1} |, \ l_j \in \calI  \label{eq:user_rate}
\end{align}
where $\bR_{l_j}$ and $\bQ_{l_j}$ are the desired signal and interference-plus-noise (I+N) covariance matrices for user $j$, at BS $l$, respectively, and are given by,
\begin{align*}
\bR_{l_j} &=  \bH_{l, l_j} \bV_{l_j} \bV_{l_j}^\dagger \bH_{l , l_j}^\dagger , l_j \in \calI  \\
\bQ_{l_j} &= \sum_{i=1}^L \sum_{k=1}^K \bH_{l, i_k}\bV_{i_k} \bV_{i_k}^\dagger \bH_{l,i_k}^\dagger + \sigma_l^2 \pmb{I}_N - \bR_{l_j},  l_j \in \calI .
\end{align*}  
We let 
\begin{align*}
 \bar{\bR}_{i_k} &=  \bH_{i , i_k}^\dagger \bU_{i_k} \bU_{i_k}^\dagger \bH_{i, i_k} , \  i_k \in \calI  \\
 \bar{\bQ}_{i_k} &= \sum_{l=1}^L \sum_{j=1}^K \bH_{l , i_k}^\dagger \bU_{l_j}  \bU_{l_j}^\dagger \bH_{l, i_k} + \bar{\sigma}_{i_k}^2 \pmb{I}_M - \bar{\bR}_{i_k}, i_k \in \calI
\end{align*}
denote the desired and I+N covariance matrices of user $i_k$, in the backwark DL network (where $ \bar{\sigma}_{i_k}^2 $ is the noise power at receiver $i_k \in \calI $).  
Moreover, we let $\bL_{l_j} \bL_{l_j}^\dagger$ be the Cholesky Decomposition of $\bQ_{l_j}$, and $\bK_{i_k} \bK_{i_k}^\dagger$ as that of $\bar{\bQ}_{i_k}$. 

We aim at maximizing the sum-rate, i.e.,
\begin{align}
(P)
\begin{cases} 
         \underset{ \lbrace \bU_{l_j} , \bV_{l_j} \rbrace}{\max}  \sum_{l_j \in \calI } ~ r_{l_j}  \\
         \st  \ \Vert \bU_{l_j} \Vert_F^2 = P_r \ ,   \Vert \bV_{l_j} \Vert_F^2 = P_t \ ,   \forall \ l_j \in \calI
\end{cases}
\end{align} 
While distributed multi-user multi-cell optimization generally entails a sum-power constraint on the users of a cell (e.g. W-MMSE~\cite{shi_wmmse_2011}), we adopt an equal power allocation among the all users within a cell: The BS power is equally split among all its UL (or DL) users, to simplify the presentation. Note that this does not affect the generality of the results.

\section{Proposed Approach}
Sum-rate maximization problems, such as $(P)$, are known to be \emph{NP-hard}~\cite{Razaviyayn_dof_2011}.  
Our proposed approach is based on a tractable lower bound formulation that transfers the sum-rate maximization, which is an originally coupled problem, into separable subproblems.    
\subsection{Problem Formulation}
We focus on the interference-limited regime (in a dense deployment for instance), where we~assume 
\begin{align}  \label{eq:intf_limited} 
&\lambda_i[\bU_{l_j}^\dagger \bQ_{l_j} \bU_{l_j}] \rightarrow \infty , \  \forall i \in \lrb{d} 
\end{align}

\begin{proposition}\label{prop:dlt_bound}
Under the conditions in~\eqref{eq:intf_limited}, the $r_{l_j}$ in~\eqref{eq:user_rate} satisfies
\begin{align}	
r_{l_j} 
& \geq \log_2 | \bI_d +  \bU_{l_j}^\dagger \bR_{l_j} \bU_{l_j} |  - \tr(  \bU_{l_j}^\dagger \bQ_{l_j} \bU_{l_j} )  \triangleq   r_{l_j}^{(LB)},  \label{eq:DLT_lb} 
\end{align}
where the gap $\Delta_{l_j} \triangleq r_{l_j}-r_{l_j}^{(LB)}$ is characterized by   
\begin{align}
\Delta_{l_j}  
&=  \tr(  \bU_{l_j}^\dagger \bQ_{l_j} \bU_{l_j} ) - \log_2|\bU_{l_j}^\dagger \bQ_{l_j} \bU_{l_j} |  \nonumber \\
&~~+ \calO(\tr[(\bU_{l_j}^\dagger \bQ_{l_j} \bU_{l_j})(\bU_{l_j}^\dagger \bR_{l_j} \bU_{l_j})^{-1}  ]) ,  \ \forall l_j \in \calI  .
\end{align} 
\end{proposition}
\begin{IEEEproof}  Refer to \cite{Ghauch_MAXSEP_16}[Appendix B]
\end{IEEEproof}
In what follows, we shall dub the quantity $r_{l_j}^{(LB)}$ Difference of Log-Trace (DLT). 
The DLT becomes significant when it is used as an alternative objective of the sum-rate objective in $(P)$.   
Note that DLT is a lower bound on the sum-rate, $R_{\Sigma}$, and can be written in the following ways:  
\begin{align}
R_{\Sigma}^{(LB)} 
&= \sum_{l_j \in \calI} \log_2 | \pmb{I}_d +  \bU_{l_j}^\dagger \bR_{l_j} \bU_{l_j} |  - \tr(  \bU_{l_j}^\dagger \bQ_{l_j} \bU_{l_j} )  \label{eq:rate} \\
&= \sum_{i_k \in \calI} \log_2 | \pmb{I}_d + \bV_{i_k}^\dagger \bar{\bR}_{i_k} \bV_{i_k} |  - \tr( \bV_{i_k}^\dagger \bar{\bQ}_{i_k} \bV_{i_k} ) \label{eq:raterev}
\end{align}
The above expressions reveal that DLT makes both the receive filters in~\eqref{eq:rate} and the transmit filters  in~\eqref{eq:raterev} decoupled, and facilitates the aimed distributed F-B implementation.  
We formulate the maximal DLT (max-DLT) criterion as a surrogate objective to the sum-rate maximization in $(P)$,
\begin{align} 
(P_{LB})
\begin{cases} 
\underset{ \lrb{\bV_{l_j},\bU_{l_j}} }{\max}  R_{\Sigma}^{(LB)}  \\
\st \ \Vert \bU_{l_j} \Vert_F^2 = P_r , \  \Vert \bV_{l_j} \Vert_F^2 = P_t , \ \forall l_j \in \calI . 
\end{cases} 
\end{align}
It should be noted that although DLT allows distributed F-B  implementation, the problem in $(P_{LB})$  is not directly solvable since it is non-convex due to the coupling between the transmit and receive filters, and the quadratic equality constraints. 

\subsection{Proposed Algorithm}
We underline that a coupled optimization like~$(P_{LB})$ can ideally be handled by a Block Coordinate Descent (BCD) approach. 
This means if the superscript $^{(n)}$ is adopted to denote the iteration number,  problem~$(P_{LB})$ is decomposed into a sequence of subproblems that are solved via F-B iterations as below 
\begin{align} \label{opt:bcd}
\!\!\!\! (J1)~& \lrb{\bU_{l_j}^{(n+1)}} \triangleq \underset{ \lrb{ \bU_{l_j} } }{\argmax} \ R_{\Sigma}^{(LB)} \left( \lrb{\bU_{l_j}} , \lrb{\bV_{l_j}^{(n)}} \right) \nonumber \\
\!\!\!\! (J2)~& \lrb{ \bV_{l_j}^{(n+1)} } \triangleq \underset{ \lrb{ \bV_{l_j} } }{\argmax} \ R_{\Sigma}^{(LB)} \left(\lrb{\bU_{l_j}^{(n+1)}}, \  \lrb{\bV_{l_j}} \right),   
\end{align}
for $n = 1,2,3 ...$. 
As seen from~\eqref{eq:rate}, at each iteration $n$, given the fixed $\lrb{\bV_{l_j}^{(n)}}$,  problem (J1) is decouples in the receive filters $\lrb{\bU_{l_j}}$, yielding  
\begin{align} 
\!\! (J1) \ 
\begin{cases}  \label{opt:rlbrx}
\underset{ \bU_{l_j} }{\min} \ \tr(  \bU_{l_j}^\dagger \bQ_{l_j} \bU_{l_j} ) -  \log_2 | \pmb{I}_d +  \bU_{l_j}^\dagger \bR_{l_j} \bU_{l_j} |   \\
\st \ \Vert \bU_{l_j} \Vert_F^2 = P_r .
\end{cases} 
\end{align}
Likewise, if the receive filters are fixed,  problem (J2) decouples, as seen from \eqref{eq:raterev}, in the transmit filters, resulting in  
\begin{align} 
\!\!\!\!\!  (J2) \
\begin{cases}  \label{opt:rlbtx}
\underset{ \bV_{i_k} }{\min} \ \tr( \bV_{i_k}^\dagger \bar{\bQ}_{i_k} \bV_{i_k} ) -  \log_2 | \pmb{I}_d + \bV_{i_k}^\dagger \bar{\bR}_{i_k} \bV_{i_k} |   \\
\st \ \Vert \bV_{i_k} \Vert_F^2 = P_t.
\end{cases} 
\end{align}
As aforementioned, the feasible sets of \eqref{opt:rlbrx} and \eqref{opt:rlbtx} are non-convex.  
 Nevertheless, we show in the result below, that their globally optimal solutions can still be found.
\begin{lemma}  \label{lem_dlt} 
	Non-homogeneous Waterfilling. \\*
	Consider the following problem,
	\begin{align} 
	\!\!\!\! \begin{cases}  
	\underset{\bX \in \mathbb{C}^{n \times r} }{\min} \ f(\bX) \triangleq  \tr( \bX^\dagger \bQ \bX ) \!-\! \log_2 | \pmb{I}_d \!+\! \bX^\dagger \bR \bX |   \\
	\st \ \Vert \bX \Vert_F^2 = \zeta . \label{opt:logtr}
	\end{cases} 
	\end{align}
	where $ \bQ \succ \pmb{0} $ and $\bR \succeq \pmb{0}, \ r < n$. Let $\bQ \triangleq \bL\bL^\dagger$ be the Cholesky factorization of $\bQ$, and $\bM \triangleq \bL^{-1} \bR \bL^{-\dagger}, \ \bM \succeq \pmb{0} $, and define the following,
	$\lrb{ \alpha_i \triangleq \lambda_i[\bM] }_{i=1}^r $ , $ \ \bPsi \triangleq v_{1:r}[\bM]$, $ \lrb{ \beta_i \triangleq \bPsi_{(i)}^\dagger (\bL^\dagger \bL)^{-1}  \bPsi_{(i)} }_{i=1}^r $. 
	Then the optimal solution for \eqref{opt:logtr} is, 
	\begin{align} \label{eq:nhwf}
	\bX^\star = \bL^{-\dagger} \bPsi \bSig^\star ,
	\end{align}
	where $\bSig^\star$ (diagonal) is the optimal power allocation. 
	Moreover, optimal power allocation in $\bSig^\star$ is, 
	\begin{align} \label{eq:optpow}
	\bSig_{(i,i)}^\star = \sqrt{\Big( 1/(1+\mu^\star \beta_i) - 1/\alpha_i  \Big)^+}, \forall i ,
	\end{align}
	where $\mu^\star$ is the unique root to $g(\mu) \triangleq \sum_{i=1}^r \beta_i \big( 1/(1+\mu \beta_i) - 1/\alpha_i  \big)^+  - \zeta$, on the interval $[ -1/(\max_i \beta_i ) , \ \infty ] $, and $g(\mu)$ is monotonically decreasing on that interval. 

\end{lemma}
\begin{IEEEproof} Refer to~\cite{Ghauch_MAXSEP_16}[Appendix C] \end{IEEEproof}

With Lemma \ref{lem_dlt}, the optimal transmit and receive filter updates are formulated as below  
\begin{align}
\!\!\! \bU_{l_j}^\star &=  \bL_{l_j}^{-\dagger} \bPsi_{l_j} \  \bSig_{l_j}^\star , \ \   \bPsi_{l_j} \triangleq v_{1:d} [ \bL_{l_j}^{-1} \bR_{l_j} \bL_{l_j}^{-\dagger} ], \forall \ l_j \ , \nonumber \\  
\!\!\! \bV_{i_k}^\star &=  \bK_{i_k}^{-\dagger} \bTh_{i_k} \  \bLam_{i_k}^\star , \ \   \bTh_{i_k} \triangleq v_{1:d} [ \bK_{i_k}^{-1} \bar{\bR}_{i_k} \bK_{i_k}^{-\dagger} ], \forall \ i_k \ , \label{eq:upt2}
\end{align}
where $ \bSig_{l_j}^\star$ and $ \bLam_{i_k}^\star$ are the optimal power allocation, given in Lemma~\ref{lem_dlt}. 
Denoting by $T$ the predefined number of F-B iterations, the max-DLT algorithm is in Algorithm~\ref{alg2}. 
\begin{algorithm}[t]
	\caption{Maximal DLT (max-DLT)} \label{alg2}
	\begin{algorithmic}
		\For{$t=1,2,..., T$}
		\State // \emph{forward network optimization: receive filter update}
		\State \hspace{.2cm} Estimate   $\bR_{l_j}, \bQ_{l_j}$, and compute $\bL_{l_j}$,    $\forall l_j$
		\State \hspace{.2cm} $\bU_{l_j} \leftarrow  \bL_{l_j}^{-\dagger}  v_{1:d} [ \bL_{l_j}^{-1} \bR_{l_j} \bL_{l_j}^{-\dagger} ] \bSig_{l_j},  \ \ \forall l_j$
		\State // \emph{backwark network optimization: transmit filter update}
		\State \hspace{.2cm} Estimate   $\bar{\bR}_{i_k}, \bar{\bQ}_{i_k}$, and compute $\bK_{i_k}$ ,  $\forall i_k$
		\State \hspace{.2cm} $\bV_{i_k} \leftarrow  \bK_{i_k}^{-\dagger}  v_{1:d} [ \bK_{i_k}^{-1} \bar{\bR}_{i_k} \bK_{i_k}^{-\dagger} ] \bLam_{i_k}, \ \ \forall i_k$
		\EndFor 
	\end{algorithmic} 
\end{algorithm} 

\paragraph*{Discussions} \label{sec:disc_NHW}
Based on the generalized eigenvalue analysis, we have that $\lrb{ \alpha_i \triangleq \lambda_i[\bL^{-1} \bR \bL^{-\dagger}] }_{i=1}^r $ are also the eigenvalues of $\bQ^{-1} \bR $. 
This means $\lrb{ \alpha_i  }$ can be viewed as a (quasi)-SINR  measure of each data stream.
The proposed method in~\eqref{eq:optpow} allocates no power to streams that have low-SINR, since $\bSig_{(i,i)}^\star$  tends to zero as $\alpha_i \rightarrow 0$. 
Moreover, as seen from~\eqref{eq:optpow}, $\lrb{\beta_i}$ models the price of activating each of the streams, mimicking the original waterfilling principle. 
The difference however is that~\eqref{eq:nhwf} fills the power level based on the SINR and cost for the stream activation, namely the non-homogeneous waterfilling solution. 
This readily enables the algorithm to not allocate power to some low SINR streams. 
Finally, since the global optimizer is found at each iteration in Algorithm \ref{alg2}, we can  conclude that  $R_{\Sigma}^{(LB)}( \lrb{\bU_{l_j}^{(n)}}, \lrb{ \bV_{l_j}^{(n)} })$ in $(P_{LB})$ is monotonically decreasing with $n$, and converges to a stationary point of the DLT~bound. 
While the `stream-control' greatly speeds up the convergence, it evidently raises fairness issues, as some users/streams with low-SINR, may not get served. This can be remedied by introducing user weights in $(P)$, with minor modifications in the problem/solutions. 

\section{Practical Aspects}
\subsection{Comparisons } \label{sec:benchmark}
Our approach is applicable to other communication scenarios such as the MIMO Interfering Broadcast Channel (MIMO IBC), the MIMO Interference Channel (MIMO IFC). We benchmark our algorithms against widely adopted ones,
\begin{itemize}
\item[o] \emph{max-SINR}~\cite{gomadam_distributed_2011} in the MIMO IMAC / MIMO IFC~/~MIMO IBC
\item[o] \emph{MMSE and Weighted-MMSE}~\cite{peters_cooperative_2011, shi_wmmse_2011} in the MIMO IFC / MIMO IBC 
\item[o] \emph{Uncoordinated} (Eigen-beamforming): each transmit (resp. receive) filter uses right (resp. let) singular eigenvectors of the desired channel
\end{itemize}
We also include relevant fast-converging algorithms,
\begin{itemize}
\item[o] \emph{CCP-WMMSE}~\cite{Nguyen_WSR_14}: an accelerated version of WMMSE algorithm for the MIMO IMAC 
\item[o] \emph{IWU}~\cite{Ghauch_IWU_15}: a fast-convergent leakage minimization algorithm for the MIMO IFC
\item[o] \emph{AIMS}: our previously proposed generalization of max-SINR~\cite{Ghauch_IWU_15}, for MIMO IMAC / MIMO IFC / MIMO IBC
\end{itemize}
Algorithms such as IWU and CCP-WMMSE use so-called \emph{turbo iterations}, where $I$ inner-loop iterations are performed within each F-B iteration. Unlike IWU where the turbo iterations are done at the BS/user (i.e., offline), these iterations are carried over-the-air for CCP-WMMSE.

\subsection{Communication Overhead} \label{sec:overhead}
The operation of the proposed scheme hinges on each transmitter and receiver's having knowledge of effective channels, for the desired and interfering links. 
We note that investigating different mechanisms for the distributed acquisition of CSI is outside the scope of the current work (we refer the reader to~\cite{Brandt_DistCSI_15}). However, we have outlined a simple mechanism that goes hand-in-hand with F-B iterations, in Fig.~\ref{fig:frame}. We recall that $T$ F-B iterations are~carried~out. 
\begin{table}
  \scriptsize
  \begin{tabular}{  c c c c c c c} 
      Pilots  & Estim. cov.  & Optimize  & ~~~Pilots   & Estim. cov.  & Optimize & \textbf{Data} \\
      ~~~  & matrices & receive  & ~~~   & matrices  & transmit \\                  
       ~~~  &  at receiver & filter & ~~~   & at transmitter & filter 
\end{tabular}
\vspace{1mm}
  \caption{ Basic structure of Forward-Backward Iteration }   
  \label{fig:frame}
  \normalsize
\end{table}


It becomes clear that each F-B iteration has an associated \emph{communication overhead}. While total overhead comprises of  bidirectional transmission of pilots, synchronization, frequency offset calibration, etc, it is dominated by the pilot overhead, if the case of cellular coordination~\cite{ayach_overhead_12}. Thus, we can safely  approximate the communication overhead by the total number of pilot symbols, for channel estimation, after $T$ F-B iterations. 
In conventional coordination, it is typical to assume $T=100 \sim 1000$ until convergence, even for small systems~\cite{Schmidt_comparison_13}. Moreover, this number increases with more BSs, cells and transmit/receive antennas, all of which are prevalent in sub-$28$ GHz systems. 
This limitation is compounded by the naturally lower coherence time of mmWave channels, thus further restricting the possible number of F-B iteration (before the channel changes). 
Indeed, simple calculations reveal that conventional algorithms would fail in these systems, as the overhead would destroy the sum-rate gains from coordination. 
Thus, we aggressively limit the number of F-B iteration to $T \leq 5$, thereby resulting in a drastic tenfold reduction in the communication overhead.       

For simplicity, we additionally assume that the minimal number of orthogonal pilots is used, i.e. $d$ pilot symbols for each UL/DL effective channel, resulting in a total of $KLd$ orthogonal pilots for each UL/DL phase.  
The total overhead for max-DLT, in the number of channel uses (c.u.), is~given~ by,  
\begin{align*}
\Omega_{\textrm{prop}} = T( \underbrace{KLd}_{  UL } + \underbrace{KLd}_{ DL } ) = 2TKLd 	
\end{align*}
The overhead is the same for schemes such as max-SINR, IWU and MMSE.  
Similar calculations can be made to estimate the overhead of CCP-WMMSE and WMMSE (in c.u.),  
\begin{align*}
&\Omega_{\textrm{ccp-wmmse}} = T[(\underbrace{KLM}_{ \substack{UL \ chann. \\ estim } }) \underbrace{\times (L-1)}_{ \substack{CSI \\ sharing}} +  \underbrace{I}_{turbo} \times ( \underbrace{KLN}_{\substack{cov. \ mat \\ upd.}}) ]    \\ 
&\Omega_{\textrm{w-mmse}} = T( \underbrace{KLd}_{ UL} + \underbrace{KLM}_{ weights} + \underbrace{KLd}_{ DL})    
\end{align*}
where $I$ denotes the number of turbo iterations. 
These simple calculation reveal that the overhead for W-MMSE and CCP-WMMSE is significantly higher than that of max-DLT. Furthermore, the turbo iteration in CCP-WMMSE (outlined in Sec~\ref{sec:benchmark}) is carried over-the-air, and thus induces a massively higher overhead, compared to other schemes. We include the overhead of these algorithms in the simulation results. 

\subsection{Complexity}
We can approximate the computational complexity of max-DLT, by noticing that it is dominated by the complexity of the Cholesky Decomposition of the I+N covariance matrix, $\mathcal{O}(N^3)$, and that of Eigenvalue Decomposition of $\bM$,  $\mathcal{O}(M^3)$, 
\begin{align*}
C_{\textup{prop}} = \calO( (M+N)^3) \ .
\end{align*}
One can verify that the above also holds for max-SINR, IWU, MMSE, and WMMSE. 
Unlike other methods, the acceleration does not require gradient/Hessian, and thus comes at a negligible added computational cost, compared to conventional algorithms.  
However, each turbo iteration for CCP-WMMSE involves running a series of semidefinite programs (using interior point solvers), which render the algorithm very costly.  

\section{Numerical Results}
\subsection{Performance in Sub-6 GHz systems } 
We start with presenting results for conventional multi-cell multi-user MIMO, to illustrate desired features of max-DLT. We refer the reader to~\cite{Ghauch_MAXSEP_16} for a detailed discussion of the simulation setup. 

\subsubsection{Single-user Multi-cell MIMO Uplink}
\begin{figure}
  \center
  \includegraphics[ height=5cm, width=8.5cm ]{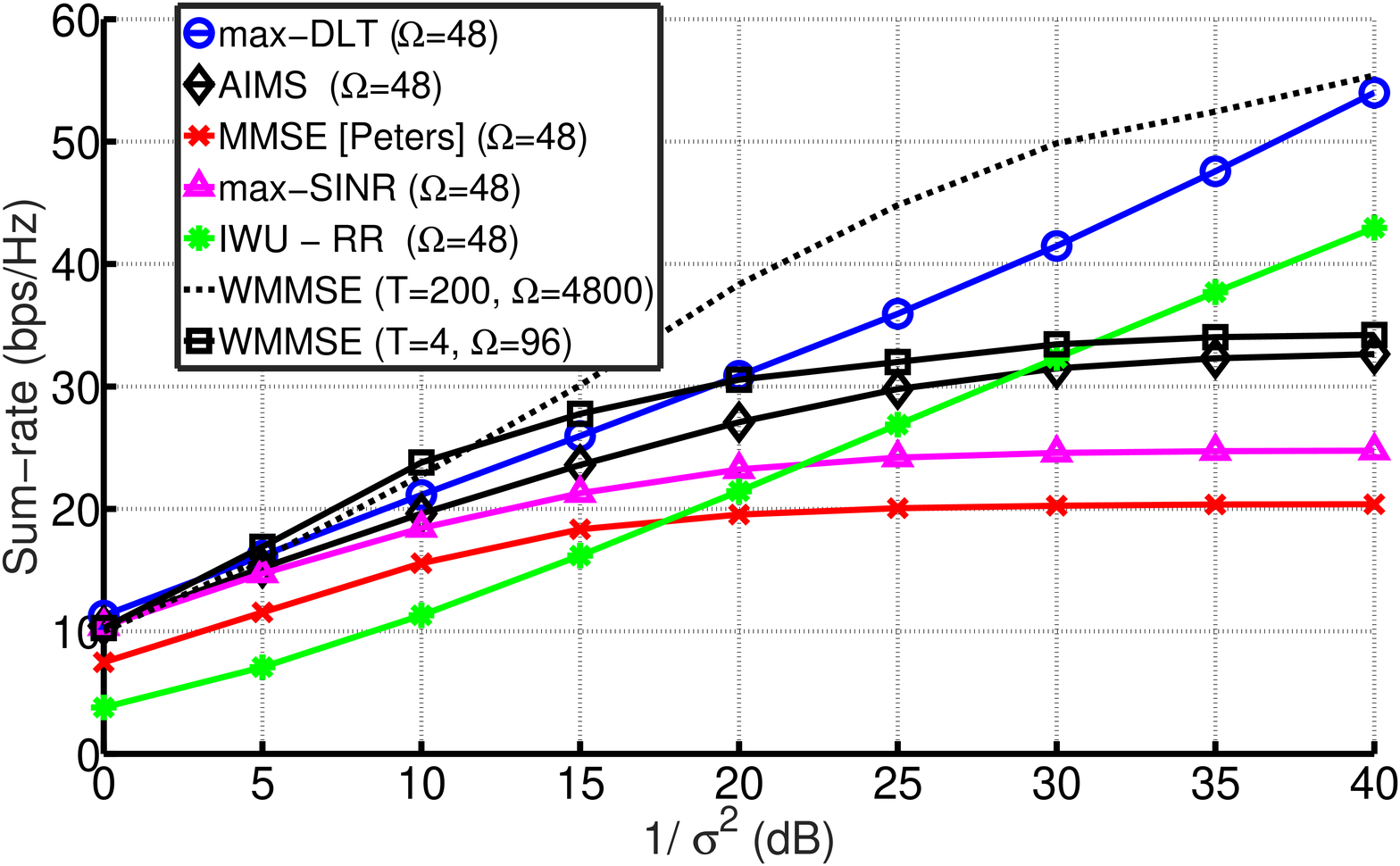}
  \caption{Ergodic sum-rate vs $1/\sigma^2$, for $L=3, K=1, M=N=4, d=2, T = 4$ (MIMO IFC)} 
  \label{fig:feasibleifc}  
  \includegraphics[trim={1.5cm 0 0 1.5cm}, clip, height=5cm, width=9.5cm ]{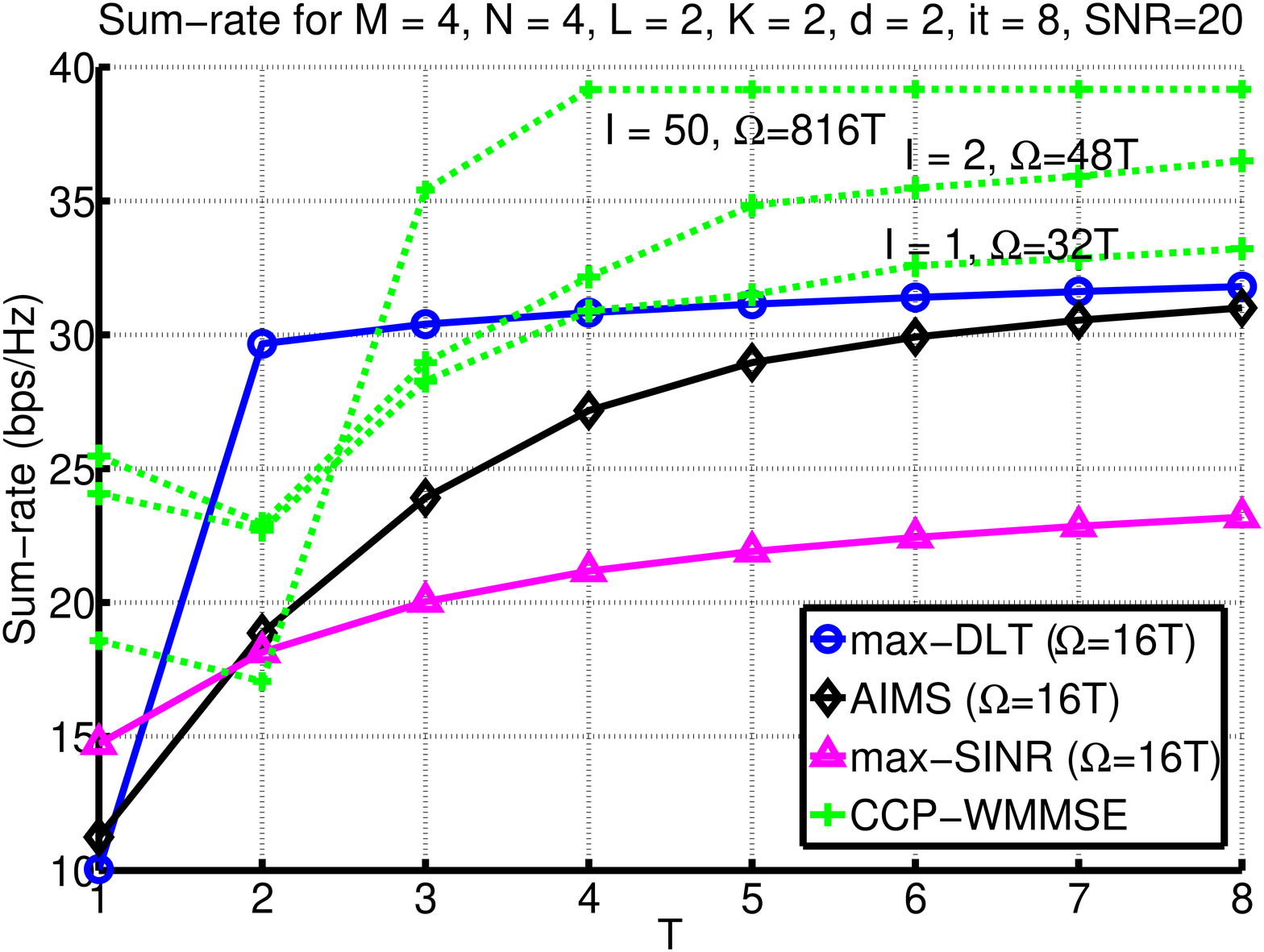}
  \caption{Ergodic sum-rate vs $T$, for $L=2, K=2, M=4, N=4, d=2$ } 
  \label{fig:SR_CCP}
  \vspace{-2em}
\end{figure}

We start with a widely used coordination test case, a MIMO IFC with $L=3, K=1, M=N=4, d=2$ where the set $T=4$ for all algorithms. 
We include W-MMSE results for  $T=4$, and $T=200$ F-B iterations (as an upper bound).  
Fig~\ref{fig:feasibleifc} reveals that while max-DLT and W-MMSE (with $T=4$) have similar performance in the low- and medium-SNR range, this gap increases sharply as the SNR increases. 
This is in spite of two-fold increase in communication overhead for WMMSE. 
Moreover, the proposed scheme yields better sum-rate performance than all benchmarks, with this gap becoming significant in the high-SNR: as the following results will show, the gap increases further with more users, antennas, and BS/cells, under a low number of F-B iterations.    
 
\subsubsection{Multi-user Multi-cell MIMO uplink}
Moving on to a larger setup with $L=2, K=2, M=4, N=4, d=2$ (MIMO IMAC), we benchmark max-DLT against the fast-converging CCP-WMMSE (Sec.~\ref{sec:benchmark}),  by varying the number of turbo iterations $I$, for CCP-WMMSE. 
Fig.~\ref{fig:SR_CCP} clearly exhibits the fast converging nature of max-DLT, that achieves $95\%$ of its final performance, after just $2$ iterations.
In the low overhead regime (for $T \leq 5$), max-DLT  outperforms CCP-WMMSE (for $I=1$), although the overhead of the latter is twice that of former. While additional turbo iterations improve slightly the CCP-WMMSE performance, the overhead increases linearly with $I$, e.g., the overhead of CCP-WMMSE with $I=2$ is threefold that of max-DLT. 
Achieving the nominal performance of CCP-WMMSE relies on convergence of the turbo iteration, which implies a (possibly arbitrary) large number of turbo iteration.  
This results in (potentially) orders-of-magnitude higher overhead/complexity (e.g. CCP-WMMSE with $I=50$). Despite its fast-converging nature, CCP-WMMSE is clearly ill-suited for the systems studied~here. 
\subsection{Performance in Dense mmWave Deployments}
Following recent measurements in the $28$ GHz band~\cite{MacCartney_PLmmW_14}, we consider a dense urban mmWave setting. The full parametrization is detailed in \cite{Ghauch_MAXSEP_16}[Sec. VI-C].     
\subsubsection{Dense Multi-user Multi-cell uplink}
\begin{figure}
	\center
	\includegraphics[ height=4cm, width=7cm]{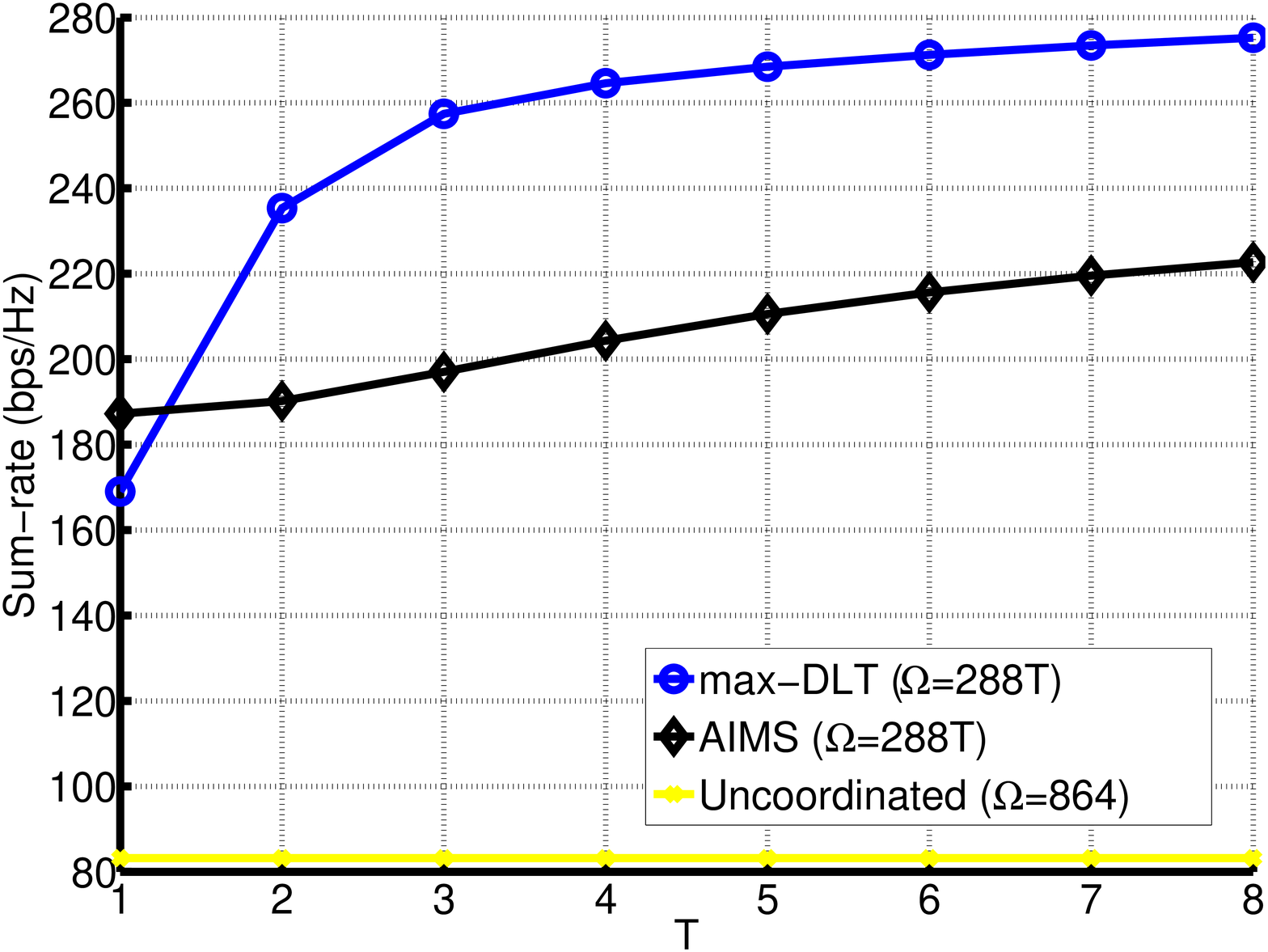}
	\caption{Ergodic sum-rate vs $T$, for dense uplink ($N=8, M=4, d=2, L=9, K=8$).  } 
	\label{fig:denseUL}
    \includegraphics[height=4cm, width=7cm]{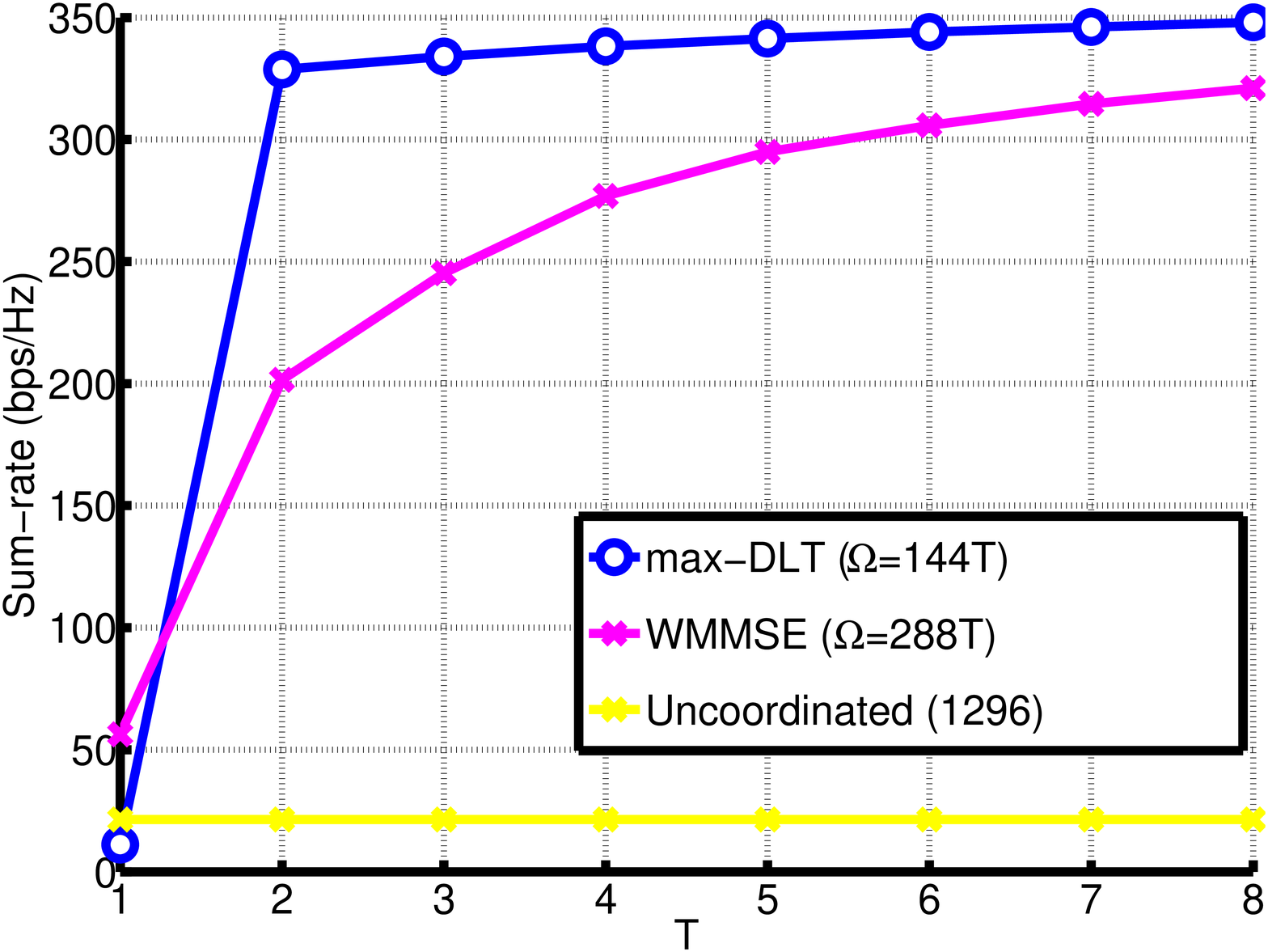}
  \caption{Ergodic sum-rate vs $T$ for dense downlink ($M=16, N=4, d=1, L=9, K=8$)  } 
  \label{fig:denseDL}
    \vspace{-2em}
\end{figure}

We consider a dense UL system with $L=9, K=8, N=8, M=4, d=2$, where the average SNR (across users) is set to $19$ dB. 
Fig.~\ref{fig:denseUL} reveals that max-DLT offers significantly better sum-rate, than all benchmarks. Interestingly, max-DLT (with $T=3$) provides a threefold increase in sum-rate with respect to the uncoordinated scheme, however, with a similar overhead. 

\subsubsection{Dense Multi-user Multi-cell downlink}
We next consider DL scenario with $L=9, K=8, M=16, N=4, d=1$, while setting the average SNR to $21$dB, and following the above simulation method. 
The fast-converging nature of max-DLT is embodied in Fig.~\ref{fig:denseDL}, where most the performance is delivered in just $2$ F-B iterations: this is due to inherent stream control feature, that allows poor quality streams to be shut down, thus converging quickly to a good sum-rate. 
Note that max-DLT assumes equal power allocation for users in each cell. In contrast, WMMSE  performs power allocation for users in each cell, as part of the algorithm.  
Despite this unfavorable setup for max-DLT, we observe a large sum-rate gain compared to WMMSE, while resulting in a $50\%$ decrease in overhead.   
Evidently, the sum-rate for WMMSE will exceed that of max-DLT, as $T$ increases (with a huge overhead). 
\subsection{Discussions}
We note the significant gap between the proposed scheme and the benchmarks, may be attributed to the fast-converging nature of the max-DLT, which is in turn due to the inherent stream-control mechanism of the non homogeneous waterfilling solution. Moreover, the drastically limited number of F-B iterations limits the performance of conventional algorithms, due to significant levels of residual interference.  
As seen in Figs.~\ref{fig:denseUL} and~\ref{fig:denseDL}, that uncoordinated transmission performs extremely poorly: max-DLT provides a threefold sum-rate improvement over uncoordinated scheme, with a similar communication overhead.  
This provides a  clear answer that low-overhead coordination is a crucial, to achieving huge sum-rate improvements in a dense multi-cell $28$ GHz mmWave system. 
This also implies that the same conclusions hold for sub-$28$ GHz systems, which are naturally more sensitive to inference.   
\section{Conclusions}
We have proposed a low-overhead algorithm for coordination, in dense multi-cell sub-$28$ GHz systems. The DLT bound - a lower bound on the sum-rate, was derived and its tightness was investigated. Moreover, we have proposed a distributed optimization algorithm (max-DLT), and showed its convergence to a stationary point of the DLT bound. 
The non-homogeneous waterfilling was derived as a solution to the optimal BS/user filter update, and its ability to turn-off low-SINR streams was underlined. 
We have tied this to the fast-convergence of the algorithm, thus enabling a tenfold reduction in communication overhead (over conventional coordination). 
Our numerical results have showed that low-overhead coordination offers huge gains, in dense sub-$28$ GHz systems.

\addcontentsline{toc}{chapter}{Bibliography}
\bibliographystyle{ieeetr}
\bibliography{ref_hadi_merged}

\begin{thebibliography}{10}

\bibitem{Andrews_5G_14}
J.~G. Andrews, S.~Buzzi, W.~Choi, S.~V. Hanly, A.~Lozano, A.~C.~K. Soong, and
  J.~C. Zhang, ``What will 5g be?,'' {\em IEEE Journal on Selected Areas in
  Communications}, vol.~32, pp.~1065--1082, June 2014.

\bibitem{Rappaprt_mmWprop}
M.~K. Samimi and T.~S. Rappaport, ``Characterization of the 28 {GHz}
  millimeter-wave dense urban channel for future {5G} mobile cellular,'' March
  2014.

\bibitem{Hur_mmWave_13}
S.~Hur, T.~Kim, D.~Love, J.~Krogmeier, T.~Thomas, and A.~Ghosh, ``Millimeter
  wave beamforming for wireless backhaul and access in small cell networks,''
  {\em IEEE Transactions on Communications,}, vol.~61, pp.~4391--4403, October
  2013.

\bibitem{Viotette_11GHz_88}
E.~J. Violette, R.~H. Espeland, R.~O. DeBolt, and F.~K. Schwering,
  ``Millimeter-wave propagation at street level in an urban environment,'' {\em
  IEEE Transactions on Geoscience and Remote Sensing}, vol.~26, pp.~368--380,
  May 1988.

\bibitem{Shokri_mmWMACSurvery_15}
H.~Shokri-Ghadikolaei, C.~Fischione, G.~Fodor, P.~Popovski, and M.~Zorzi,
  ``Millimeter wave cellular networks: A {MAC} layer perspective,'' {\em IEEE
  Transactions on Communications}, vol.~63, pp.~3437--3458, Oct 2015.

\bibitem{METISD62}
{METIS D6.2}, ``Initial report on horizontal topics, first results and {5G}
  system concept,'' March 2014.

\bibitem{gomadam_distributed_2011}
K.~Gomadam, V.~R. Cadambe, and S.~A. Jafar, ``A distributed numerical approach
  to interference alignment and applications to wireless interference
  networks,'' {\em {IEEE} Transactions on Information Theory}, vol.~57,
  pp.~3309--3322, June 2011.

\bibitem{schmidt_minimum_2009}
D.~Schmidt, C.~Shi, R.~Berry, M.~Honig, and W.~Utschick, ``Minimum mean squared
  error interference alignment,'' in {\em 2009 Conference Record of the
  Forty-Third Asilomar Conference on Signals, Systems and Computers}, pp.~1106
  --1110, Nov. 2009.

\bibitem{shi_wmmse_2011}
Q.~Shi, M.~Razaviyayn, Z.-Q. Luo, and C.~He, ``An iteratively weighted {MMSE}
  approach to distributed sum-utility maximization for a {MIMO} interfering
  broadcast channel,'' {\em IEEE Transactions on Signal Processing}, vol.~59,
  no.~9, pp.~4331--4340, 2011.

\bibitem{Schmidt_comparison_13}
D.~Schmidt, C.~Shi, R.~Berry, M.~Honig, and W.~Utschick, ``Comparison of
  distributed beamforming algorithms for {MIMO} interference networks,'' {\em
  IEEE Transactions on Signal Processing}, vol.~61, pp.~3476--3489, July 2013.

\bibitem{Komulainen_EffCSI_13}
P.~Komulainen, A.~T{\"o}lli, and M.~Juntti, ``Effective {CSI} signaling and
  decentralized beam coordination in {TDD} multi-cell {MIMO} systems,'' {\em
  IEEE Transactions on Signal Processing}, vol.~61, pp.~2204--2218, May 2013.

\bibitem{Nguyen_WSR_14}
D.~H.~N. Nguyen and T.~Le-Ngoc, ``Sum-rate maximization in the multicell {MIMO}
  multiple-access channel with interference coordination,'' {\em IEEE
  Transactions on Wireless Communications}, vol.~13, pp.~36--48, January 2014.

\bibitem{Brandt_FastConv_15}
R.~Brandt and M.~Bengtsson, ``Fast-convergent distributed coordinated precoding
  for {TDD} multicell {MIMO} systems,'' in {\em IEEE 6th International Workshop
  on Computational Advances in Multi-Sensor Adaptive Processing (CAMSAP)},
  pp.~457--460, Dec 2015.

\bibitem{Ghauch_IWU_15}
H.~Ghauch, T.~Kim, M.~Bengtsson, and M.~Skoglund, ``Distributed low-overhead
  schemes for multi-stream {MIMO} interference channels,'' {\em IEEE
  Transactions on Signal Processing}, vol.~63, pp.~1737--1749, April 2015.

\bibitem{Ghauch_MAXSEP_16}
H.~Ghauch, T.~Kim, M.~Bengtsson, and M.~Skoglund, ``Sum-rate maximization in
  sub-28-ghz millimeter-wave mimo interfering networks,'' {\em IEEE Journal on
  Selected Areas in Communications}, vol.~35, pp.~1649--1662, July 2017.

\bibitem{Razaviyayn_dof_2011}
M.~Razaviyayn, G.~Lyubeznik, and Z.-Q. Luo, ``On the degrees of freedom
  achievable through interference alignment in a {MIMO} interference channel,''
  in {\em Signal Processing Advances in Wireless Communications (SPAWC), 2011
  IEEE 12th International Workshop on}, pp.~511--515, June 2011.

\bibitem{peters_cooperative_2011}
S.~W. Peters and R.~W. Heath, ``Cooperative algorithms for {MIMO} interference
  channels,'' {\em {IEEE} Transactions on Vehicular Technology}, vol.~60,
  pp.~206--218, Jan. 2011.

\bibitem{Brandt_DistCSI_15}
R.~Brandt and M.~Bengtsson, ``Distributed {CSI} acquisition and coordinated
  precoding for {TDD} multicell {MIMO} systems,'' {\em IEEE Transactions on
  Vehicular Technology}, vol.~PP, no.~99, pp.~1--1, 2015.

\bibitem{ayach_overhead_12}
O.~El~Ayach, A.~Lozano, and R.~Heath, ``On the overhead of interference
  alignment: Training, feedback, and cooperation,'' {\em IEEE Transactions on
  Wireless Communications}, vol.~11, no.~11, pp.~4192--4203, 2012.

\bibitem{MacCartney_PLmmW_14}
G.~R. MacCartney, M.~K. Samimi, and T.~S. Rappaport, ``Omnidirectional path
  loss models in new york city at 28 {GHz} and 73 {GHz},'' in {\em 2014 IEEE
  25th Annual International Symposium on Personal, Indoor, and Mobile Radio
  Communication (PIMRC)}, pp.~227--231, Sept 2014.

\end{thebibliography}

\end{document}